\documentclass[journal,comsoc]{IEEEtran}
\makeatletter\if@twocolumn\PassOptionsToPackage{switch}{lineno}\else\fi\makeatother

\usepackage{graphicx}
\usepackage[T1]{fontenc}

\ifCLASSOPTIONcompsoc
\usepackage[nocompress]{cite}
\else
\usepackage{cite}
\fi

\usepackage[table,xcdraw]{xcolor}
\usepackage{tabularx,booktabs}
\newcolumntype{C}{>{\centering\arraybackslash}X} 
\newcolumntype{b}{X}
\newcolumntype{s}{>{\hsize=.5\hsize}X}
\newcolumntype{v}{>{\hsize=.3\hsize}X}
\usepackage{graphicx}
\usepackage[colorinlistoftodos]{todonotes}
\usepackage[colorlinks=true, allcolors=blue]{hyperref}
\usepackage{caption}
\usepackage{latexsym}
\usepackage{enumitem}
\usepackage{amssymb}
\usepackage{subcaption}
\usepackage{mathtools}
\usepackage{multirow}
\usepackage{float}
\usepackage[linesnumbered,ruled,vlined]{algorithm2e} 
\usepackage{cprotect}
\usepackage{xspace}
\usepackage{amsmath} 
\usepackage{soul}
\usepackage{hyphenat}
\usepackage{epstopdf}
\usepackage{threeparttable}
\usepackage{pifont}
\usepackage{siunitx} 
\usepackage{algorithm2e}
\def\Plus{\texttt{+}}

\DeclareMathOperator*{\argmax}{argmax}

\usepackage{tikz}
\usepackage{tkz-tab}
\usetikzlibrary{automata,arrows,positioning,calc}
\usetikzlibrary{shapes,snakes}

\def\Plus{\texttt{+}}
\def\Minus{\texttt{-}}

\makeatletter
\AtBeginDocument{\@ifpackageloaded{textcomp}{}{\usepackage{textcomp}}}
\newcommand{\removelatexerror}{\let\@latex@error\@gobble}

\makeatother

\begin{document}

	\title{\huge Online Primary Channel Selection for Dynamic Channel Bonding in High-Density WLANs}		
	
	\author{Sergio~Barrachina-Mu\~noz, Francesc Wilhelmi, and Boris~Bellalta\thanks{All the authors are with the Wireless Networking research group at Universitat Pompeu Fabra, Barcelona, Spain (e-mail: \{sergio.barrachina, francisco.wilhelmi, boris.bellalta\}@upf.edu). This  work  has  been  partially  supported  by  the  Spanish Ministry of Economy and Competitiveness under the Maria de Maeztu  Units  of  Excellence  Programme  (MDM-2015-0502),  by  a  Gift from the Cisco University Research Program (CG\#890107, Towards Deterministic Channel Access in High-Density WLANs) Fund, a corporate advised fund of Silicon Valley Community Foundation, and by PGC2018-099959-B-I00 (MCIU/AEI/FEDER,UE). The work done by S. Barrachina-Mu\~noz is supported by a FI grant from the Generalitat de Catalunya.}}

	\maketitle

	\begin{abstract}
		

		
		
		In order to dynamically adapt the transmission bandwidth in wireless local area networks (WLANs), dynamic channel bonding (DCB) was introduced in IEEE 802.11n. It has been extended since then, and it is expected to be a key element in IEEE 802.11ax and future amendments such as IEEE 802.11be.
		While DCB is proven to be a compelling mechanism by itself, its performance is deeply tied to the primary channel selection, especially in high-density (HD) deployments, where multiple nodes contend for the spectrum. Traditionally, this primary channel selection relied on picking the most free one without any further consideration.
		In this paper, in contrast, we propose dynamic-wise (DyWi), a light-weight, decentralized, online primary channel selection algorithm for DCB that maximizes the expected WLAN throughput by considering not only the occupancy of the target primary channel but also the activity of the secondary channels. Even when assuming important delay costs due to primary switching, simulation results show a significant improvement both in terms of average delay and throughput.

	\end{abstract}
	

	\IEEEpeerreviewmaketitle

	\section{Introduction}\label{sec:introduction}
	
	\IEEEPARstart{M}ODERN applications like augmented reality, virtual reality, or real-time 8K video are pushing next-generation (nextGen) wireless local area networks (WLANs) to support ever-increasing demands on performance. In addition, the characteristic nextGen high-density (HD) deployments, where numerous wireless devices will contend for accessing the medium, hinders even more the challenge of providing high throughput and low latency.
	
	In order to improve the performance of nextGen WLANs, we focus on spectrum efficiency. In particular, two well-known techniques have been widely studied in this regard: channel allocation (CA) and dynamic channel bonding (DCB). CA is the method to assign portions of the spectrum (or channels) to one or multiple WLANs. In contrast, DCB is a technique whereby two or more channels are bonded according to their instant occupancy, enabling wider bandwidths per transmission, and thus potentially reaching higher data rates.
	
	Although much has been understood from the works on CA and DCB in the literature, little has been assessed with respect to combining CA with DCB altogether in WLANs, particularly for high-density (HD) deployments. Nevertheless, while DCB has shown a tremendous potential to outperform traditional single-channel (SC) \cite{park2011ieee,deek2013intelligent,barrachina2019dynamic}, its performance is severely tied to the CA. Especially, the primary channel selection is critical since it runs the backoff procedure of the carrier sense multiple access with collision avoidance (CSMA/CA) protocol. Still, always selecting the least occupied channel as the primary is no longer appropriate for DCB since potential bonds with adjacent channels should be also considered. 
	
	
	As for works combining CA and DCB for WLANs, we find a distributed algorithm for jointly allocating channel center frequencies and bandwidths \cite{herzen2013distributed} or
	a centralized approach for maximizing the network fairness \cite{bellalta2016interactions}.
	Recently, a heuristic algorithm for primary channel selection based on the bonding direction likelihoods was presented in \cite{khairy2018renewal}. However, such likelihoods are estimated by assuming a known number of users in each channel. It is worth noticing that the aforementioned works consider fully-backlogged traffic, thus missing insights on more realistic patterns.
	Finally, an uncertain traffic CA approach was presented in \cite{nabil2017adaptive}.
	Still, a centralized controller in the backend is required.
	
	In this paper, we formulate dynamic-wise (DyWi), a decentralized, lightweight algorithm that leverages information about the sensed spectrum occupancy of the whole allocated bandwidth of a node (i.e., primary and secondary channels) in an online manner. Based on such occupancy, the primary channel is selected with the aim of maximizing the expected throughput, considering not only the activity of the target primary channel but also the potential bonds that could be established with its adjacent channels.	
	DyWi is adaptive in the sense that a new primary channel is only adopted when the WLAN performance is below a given satisfaction threshold. Besides, since DyWi relies just on local information, neither neighbor messaging nor a central controller is required. This property makes DyWi suitable to be implemented in off-the-shelf access points (APs), avoiding costly inter-WLAN collaboration. 
	
	Simulations in IEEE 802.11ax HD deployments show important improvements with respect to traditional primary selection, even when considering substantial delays due to channel switching.
	
	\section{Primary channel selection for DCB}  \label{sec:primary}
	
	\subsection{Dynamic channel bonding}
	
	DCB was first introduced in IEEE 802.11n (2009), where two contiguous 20-MHz channels could be bonded to form a single 40-MHz channel. Then, IEEE 802.11ac (2013) extended the DCB capability to bond up to 8 20-MHz channels reaching a maximum of 160-MHz bandwidth. While IEEE 802.11ax (2019) keeps such a limit, future amendments like EXtreme throughput  (i.e., IEEE 802.11be) aim to support 320-MHz transmissions.
	
\begin{figure}[t]
		\centering
		\includegraphics[width=0.48\textwidth]{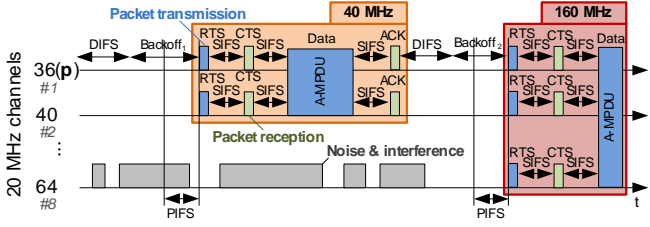}
		\caption{CSMA/CA operation of a node operating under DCB in the IEEE 802.11ac/ax channelization scheme for the UNII-1 and UNII-2 bands. The numbers preceded by a pound symbol (\#) represent a simpler channel indexation.}
		\label{fig:csma_dcb}
	\end{figure}

	Fig. \ref{fig:csma_dcb} shows the operation timeline of a node implementing DCB with primary channel $p=1$. Note that the bandwidth selection is decided according to the occupancy of the secondary channels during the PCF Interframe Space (PIFS) previous to the backoff termination. Accordingly, transmissions of 40 and 160 MHz are performed in the example after the expiration of the first and second backoff, respectively.
	
	\subsection{Online selection of the primary channel}
	
	Let an AP belonging to WLAN $w$ operate under DCB and have allocated the full available bandwidth according to a channelization scheme $\mathcal{C}$ (e.g., $|\mathcal{C}_\text{20-MHz}| = 8$ 20-MHz channels like in Fig. \ref{fig:csma_dcb}),\footnote{The channelization (or allowed transmission channels) in Fig. \ref{fig:csma_dcb} is $\mathcal{C}=\{\{1\},\{2\},...,\{8\},\{1,2\},...,\{7,8\},\{1,2,3,4\},\{5,6,7,8\},\{1,...,8\}\}$, where $\mathcal{C}_\text{20-MHz} = \{\{1\},\{2\},...,\{8\}\}\subset \mathcal{C}$ is the set of 20-MHz channels inside $\mathcal{C}$, and the rest are bonded channels.} and random primary channel $p\in\mathcal{C}_\text{20-MHz}$. Note that throughout the rest of the paper, we also use $b=20$ MHz to denote the bandwidth of a single channel (or basic channel).
	
\begin{figure}[t]
		\centering
		\includegraphics[width=0.49\textwidth]{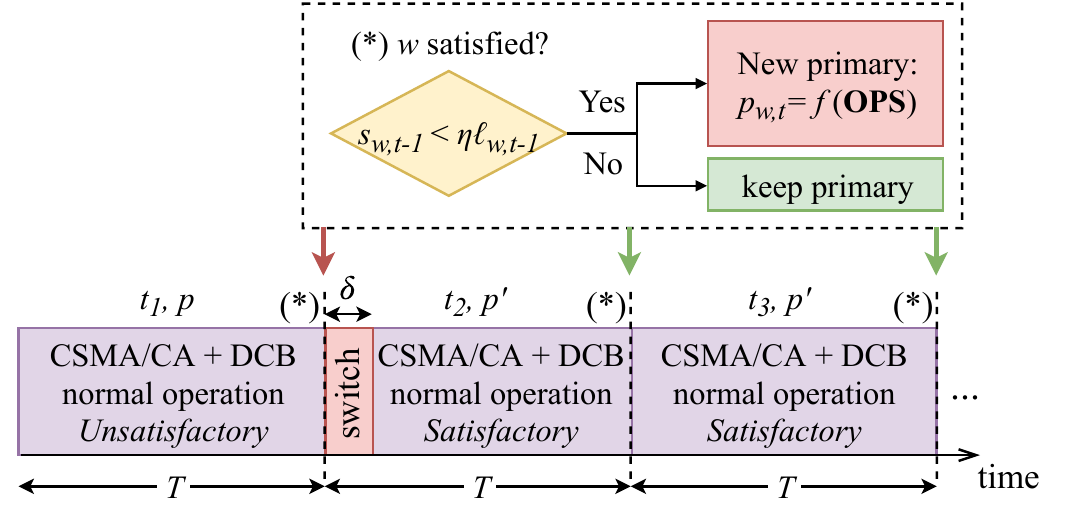}
		\caption{Example of the online primary channel selection. Since $w$ is not satisfied at the end of iteration $t_1$, the primary channel is switched from $p_{w,1}=p$ to $p_{w,2}=p'$. In contrast, with the new primary, $w$ gets satisfied in $t_2$ and keeps $p'$ in $t_3$.}
		\label{fig:dps_diagram}
	\end{figure}

	For the sake of identifying in an online manner a convenient primary channel, we rely on an iterative algorithm. In essence, $w$ periodically makes a decision about the primary channel selection, where each decision instant represents the beginning of an iteration in the online algorithm. Specifically, at the beginning of a given iteration $t$, the AP of $w$ computes the throughput achieved during the last iteration $t-1$ and acts according to a satisfaction condition.\footnote{In this work, we focus on the successful downlink traffic as the main performance metric. However, the algorithm can be easily extended to consider other parameters such as latency.}
	Namely, the primary channel remains the same if $w$ is satisfied because sufficient traffic has been successfully sent during the last iteration\footnote{We rely just on data from the last iteration for lowering memory demands and enabling fast adaptability in dynamic environments.} $t\Minus 1$, i.e., $s_{w,t\Minus 1} \geq \eta \ell_{w,t\Minus 1}$, where $\eta$ is the satisfaction ratio and $\ell_{w,t\Minus 1}$ is the actual generated traffic load in that iteration.
	Otherwise, $w$ will change its primary at the cost of remaining inactive a period $\delta$ due to factors like message broadcasting to the associated stations (STAs) or new setup configuration. The temporal evolution of the general procedure is displayed in Fig. \ref{fig:dps_diagram} through a particular example.
	
	Algorithm \ref{alg:idps} shows the pseudocode of three different online primary selection (OPS) schemes: dynamic-random (DR), dynamic-free (DF), and dynamic-wise (DyWi or DW).
	In the event of an unsatisfactory iteration, DR selects a new primary channel uniformly at random, DF picks the one found most free during the last iteration, and DyWi selects it based on the forecast throughput given the probability of bonding in every possible bandwidth.
	Such probabilities are estimated by periodically measuring the energy in all the secondary channels as done during the PIFS period.
	We use both DR and DF as baselines.
	Note that traditional fixed primary (FP) allocation does not change the primary channel under any circumstances.
	
	\begin{algorithm}[t]
		\footnotesize
		\KwIn{$p_{w,t}$, $\eta$, OPS, $\mathcal{C}$\;}
		iteration $t\leftarrow 0$\;
		\While{WLAN $w$ active}{
			
			\While{iteration $t$ not finished}{
				CSMA/CA normal operation with DCB\;
			}
			$t \leftarrow t \Plus 1$\;
			
			$s_{w,t\Minus1} \leftarrow$ \texttt{compute\_throughput($t\Minus1$)}\;
			
			$\{\boldsymbol{\rho}_{w,t\Minus1},\boldsymbol{\pi}_{w,t\Minus1}\} \leftarrow$ \texttt{get\_occupancy($t\Minus1$)}\;
			
			\If{$s_{w,t\Minus1} < \eta  \ell_{w,t\Minus 1}$}{
				\textbf{switch} (OPS)\\
				\hspace{3mm} \textbf{case} DR:
				$p_{w,t} \leftarrow U( \{1,2,\dots,|\mathcal{C}_\text{20-MHz}|\} \setminus p_{w,t\Minus1})$\;
				
				\hspace{3mm} \textbf{case} DF:
				$p_{w,t} \leftarrow \argmax\limits_{p_{w,t} \neq p_{w,t\Minus1}} \boldsymbol{\pi}_{w,t\Minus1}(p)$\;

				\hspace{3mm} \textbf{case} DW: $p_{w,t} \leftarrow \argmax\limits_{p_{w,t} \neq p_{w,t\Minus1}} \hat{r}_{w,t}$; \textcolor{gray}{\hspace{2mm}// see (\ref{eq:argmax} - \ref{eq:probs})}\\
				
				\texttt{apply\_conf($p_{w,t}$)}\;
				\texttt{wait($\delta$)}\;
				
			}
			
		}
		
		\caption{Online primary selection. OPS refers to the selected online primary selection scheme.}
		\label{alg:idps}
	\end{algorithm}
	\normalsize
	
	\subsection{Dynamic-wise (DyWi) primary channel selection}
	
	A main question arises when considering DyWi regarding the way the primary channel is selected. Assume an example scenario where a WLAN $w$ is allocated 80 MHz accounting from channel 1 to 4. At the end of iteration $t\Minus1$, $w$ is unsatisfied and must change its primary from $p_{w,t\Minus1} = 2$ to $p_{w,t}$. Assume also that the probabilities of finding free each of its allocated 20-MHz channels in iteration $t\Minus1$ was $\boldsymbol{\pi}_{w,t\Minus1} = [0.93, 0.38, 0.85, 0.85]$, where $\boldsymbol{\pi}_{w,t\Minus1}[c]$ is the probability of finding the basic channel $c$ free. Then, two main options may be contemplated as best choice for selecting $p_{w,t}$ to maximize the throughput of the upcoming iteration, $s_{w,t}$: \textit{i}) to pick the primary with highest probability to be free (i.e., $p_{w,t} = 1$ in this case), or \textit{ii}) to pick the primary providing the highest potential average data rate considering both its probability to be free, as well as the probability of the channels nearby (e.g., $p_{w,t} \in \{3,4\}$).
	
	We tackle this point at issue by proposing a maximization problem for the forecast throughput of WLAN $w$ at iteration $t$. Notice that maximizing the successful data rate is the same as maximizing the throughput. Besides, the average data rate is given by the probability of transmitting at each possible bandwidth. Then, we can formulate the problem as
	\begin{equation}	\label{eq:argmax}
	\begin{aligned}
	\argmax_{p_{w,t} \neq p_{w,t\Minus1}} & \hat{r}_{w,t}(p_{w,t}),\text{with }\\
	\hat{r}_{w,t}(p_{w,t}) & =  \sum_{n_{\text{c}}\in \mathcal{N}} \mathbb{P}_{w,t}\big(p_{w,t},n_{\text{c}}\big) r_{w,t}\big(n_\text{c}\big)\text{,}
	\end{aligned}
	\end{equation}
	where $\hat{r}_{w,t}(p_{w,t})$ is the expected data rate by WLAN $w$ at iteration $t$ for new primary $p_{w,t}$, $n_\text{c}$ is the number of bonded channels,\footnote{In the IEEE 802.11ax amendment, basic channels are $b=20$ MHz and $n_\text{c} \in \mathcal{N} = \{1,2,4,8\}$, for 20 to 160-MHz allowed bandwidths.} $\mathbb{P}_{w,t}(p_{w,t},n_\text{c})$ is the probability that $w$ transmits in $n_\text{c}$ contiguous channels in the starting iteration given $p_{w,t}$ is selected,\footnote{Note that in single-channel, $\mathbb{P}_{w,t}(n_\text{c},p) = 0$ for $\forall n_\text{c} \neq 1, \forall w,t$.} and $r_w(n_\text{c})$ is the data rate given the bandwidth $n_\text{c}b$. Note that $r_w$ also depends on the modulation coding scheme (MCS) index, which will vary according to the signal-to-noise ratio (SNR) at the STA.
	
	In order to estimate the probability of transmitting in each possible combination of channels, we rely on the empirical probability that a set of $n_\text{c}$ channels was free during the last iteration $t-1$ given the primary $p_{w,t}$ and channelization $\mathcal{C}$, $\rho_{w,t-1}(p_{w,t},n_\text{c})$, which was updated during the backoff operations in iteration $t\Minus 1$. Essentially,
	\begin{equation}
	\rho_{w,t-1}(p_{w,t}, n_c) = \mathbb{E}_{t-1}\big(P_\text{rx}(c) < \text{CCA}, \forall c \in \mathcal{C}_\text{tx}(p_{w,t},n_c)\big)\text{,}
	\end{equation}
	where $P_\text{rx}(c)$ is the power received at the basic channel $c$, CCA is the clear channel assessment (CCA) threshold, and $\mathcal{C}_\text{tx}(p_{w,t},n_c)$ is the set of basic channels used in the transmission, which is mandated by the channelization scheme $\mathcal{C}$. For instance, following the IEEE 802.11ac/ax channelization, for primary $p=6$ and $n_\text{c} =2$, the corresponding 40-MHz bonded channel is given by $\mathcal{C}_\text{tx}(6,2) = \{5,6\}$. 
	
	Since DCB is implemented, the largest available bandwidth is always picked per transmission. Hence, the probability of transmitting in a certain bandwidth is contingent on the probability of transmitting in higher bandwidths. Specifically,
	\begin{equation} \label{eq:probs}
	\mathbb{P}_{w,t}(p_{w,t}, n_\text{c}) = \rho_{w,t\Minus1}(p_{w,t}, n_\text{c}) - \sum_{n_\text{c}' \in \{\mathcal{N} | n_\text{c}' > n_\text{c}\}} \mathbb{P}_{w,t}(p_{w,t},n_\text{c})\text{,}
	\end{equation}
	where the expression $n_\text{c}' \in \{\mathcal{N} | n_\text{c}' > n_\text{c}\}$ is a constraint for subtracting the probability of transmitting in wider bandwidths. For instance, following again the IEEE 802.11ax channelization scheme for both the UNII-1 and UNII-2 bands (i.e., $\mathcal{C}_\text{20-MHz} = \{1,...,8\}$), we define $\mathbb{P}_{w,t}(p,8) = \rho_{w,t-1}(p,8)$ for $n_\text{c}=8$ (i.e., 160-MHz). Similarly, on the other end, $\mathbb{P}_{w,t}(p, 1) = \rho_{w, t-1}(p,1) - \mathbb{P}_{w,t}(p, 2) - \mathbb{P}_{w,t}(p, 4) - \mathbb{P}_{w,t}(p, 8)$ for $n_\text{c}=1$ (i.e., 20-MHz).
	
	As for the complexity of the presented iterative algorithms, note that they are computational lightweight; especially DR since it does not keep track of any data. Despite DF's complexity increases with the number of channels, $\mathcal{O}_\text{DF}\big(|\mathcal{C}_\text{20-MHz}|\big)$, it is also low. DyWi's complexity, however, is bounded by $\mathcal{O}_\text{DW}\big(|\mathcal{C}_\text{20-MHz}| (\log_{2}|\mathcal{C}_\text{20-MHz}| + 1)^2\big)$ for the IEEE 802.11ax channelization. Nonetheless, DyWi is completely tractable in operation time by off-the-shelf network cards since the number of possible bonds in the 5-GHz band is still small.
	
	\section{System model}
	
	For evaluating the performance of the algorithms presented in Section \ref{sec:primary}, we simulate IEEE 802.11ax HD deployments using the Komondor \cite{barrachina2019komondor} wireless network simulator v1.2.1c.
	For simplicity, we consider negligible propagation delay, downlink traffic, and WLANs composed by one AP and one STA. The packet arrival process at each AP follows a Poisson process generating packets every $t_n \sim \text{Exponential}\big(L_\text{d} / \bar{\ell}\big)$, where $L_\text{d} / \bar{\ell}$ is the average packet arrival rate given a fixed packet length $L_\text{d}$ and average arrival bit rate $\bar{\ell}$ \cite{barrachina2019overlap}.
	
	As for the packet reception model, we consider that a packet is lost if the signal-to-interference-plus-noise ratio (SINR) perceived at the receiver does not accomplish the capture effect (CE). Note that transmitted power is spread over the channels used in the transmission bandwidth.
	We also consider the same CCA (-82 dBm) both in primary and secondary channels to make channel access more restrictive, as proposed in \cite{park2011ieee}. 

	\begin{table}
		\caption{Evaluation setup.}
		\label{table:appendix_table}
		\centering
		\footnotesize
		\begin{tabularx}{.48\textwidth}{ccc}
			\toprule
			\textbf{Parameter}     & \textbf{Description}              & \textbf{Value} \\ 
			\midrule
			$f_\text{c}$ & Central frequency           & 5.25 GHz  \\
			$b$ & Basic channel bandwidth          & 20 MHz \\
			$L_\text{d}$       & Data packet size           & 12000 bits     \\ 
			$N_\text{b}$		& Buffer capacity & 150 packets \\
			$N_\text{a}$       & Max. no. of aggregated packets in a frame & 64             \\  
			$\text{CW}_\text{min}$ & Min. contention window            & 16             \\ 
			$m$                    & No. of backoff stages          & 5              \\   
			MCS						& IEEE 802.11ax MCS index							& 0 - 11		\\
			$\eta$                 & MCS's packet error rate        & 0.1           \\ 
			CCA                    & CCA threshold                               & -82 dBm        \\ 
			$P_\text{tx}$          & Transmission power                & 15 dBm         \\ 
			$G_\text{tx}$         & Transmitting gain                 & 0 dB           \\ 
			$G_\text{rx}$         & Reception gain                    & 0 dB           \\ 
			$\text{PL}(d)$		& \begin{tabular}[c]{@{}c@{}}TMB indoor path loss for 11ax\end{tabular} 			& see \cite{adame2019tmb}		\\
			CE                     & Capture effect threshold          & 20 dB          \\ 
			$N$                      & Background noise level            & -95 dBm        \\
			\midrule
			$\mathcal{C}$ & Channelization for UNII-1 \& UNII-2          & 36(1) - 64(8) \\
			$|\mathcal{C}_\text{20-MHz}|$ & No. of 20-MHz channels in the system        & 8 \\
			OPS & Online primary selection scheme        & DR, DF, DW \\
			$T$                   & Iteration duration                      & 1 \si{s}\\
			$T_\text{obs}$                   & Simulation duration                      & 25 \si{s}\\
			$\eta$                   & Satisfaction ratio                      & 0.9\\
			$\delta$       & Switching delay cost                & 0, 100 \si{ms}         \\
			
			\bottomrule
			
		\end{tabularx}
	\end{table}
	
	\begin{figure}[t]
		\centering
		\includegraphics[width=0.4\textwidth]{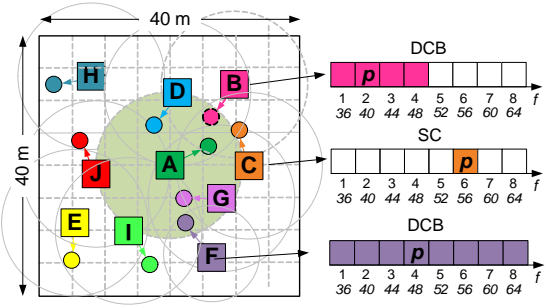}
		\caption{Deployment with WLAN A in the center. The IEEE 802.11ac/ax 20-MHz channel indexation is displayed in italic. WLANs may implement SC or DCB in different allocated bandwidths.}
		\label{fig:map_central}
	\end{figure}
	
	In this Section, we contemplate a dense 40x40 \si{m}$^2$ deployment like the one shown in Fig. \ref{fig:map_central}. One WLAN (A) remains located at the center for every scenario, and other 9 WLANs are spread uniformly at random in the area. The only condition is that any pair of APs must be separated at least $d_{\text{AP-AP}}^\text{min} = 10$ m. The STA
	of each WLAN is located also uniformly at random at a distance $d_\text{AP-STA} \in [d_\text{AP-STA}^\text{min}, d_\text{AP-STA}^\text{max}] = [1, 5]$ m from the AP. Note that WLANs are not required to be within the carrier sense range of the others, i.e., the simulations capture spatial distribution effects.
	
	Regarding the CA, all the WLANs are set with a random primary channel in the eight basic channels considered in the system (i.e., $p_\text{w} \sim U[1,8], \forall w$). The set of allocated basic channels is assigned uniformly at random as well. That is, the number of allowed basic channels for transmitting in $w$ is $|\mathcal{C}_{\text{tx},w}| \sim U\{1,2,4,8\}, \forall w$, except for WLAN A, which is allocated the whole bandwidth channel in all the scenarios (i.e., $\mathcal{C}_{\text{tx},A} = \{1,...,8\}$).		
	While the DCB capabilities of the rest of WLANs are also set uniformly at random (i.e., they implement SC or DCB with same probability 1/2), A is fixed to DCB. 
	
	We generate $N_\text{D} = 200$ random deployments following the aforementioned conditions and evaluate $N_\ell = 17$ values of A's average traffic load $\bar{\ell}_A$ ranging from 1 to 400 Mbps. The rest of WLANs are set with random average traffic load inside this range, i.e., $\bar{\ell}_w \sim U[1,400]$ Mbps. In addition to traditional FP, we consider the $N_\text{P} = 3$ OPS schemes proposed in Section \ref{sec:primary} (i.e., DR, DF, DW). For the latter ones, two switching delay costs $\delta \in \{0, 100\}$ ms are assessed. Consequently, we simulate $N_\text{D} \times N_\ell \times (1 + 2N_\text{P}) = 23,800$ scenarios. The simulation time of each scenario is $T_\text{obs}=25$ seconds. As for the configuration of the online algorithms, we set the iteration time $T = 1$ \si{s} and satisfaction ratio $\eta = 0.9$. Note that we consider a value of $\eta$ smaller than 1 to provide stability to the algorithm.
	
	\section{Performance Evaluation}
	
	\begin{figure}[t]
		
		\centering
		\begin{subfigure}{0.24\textwidth}
			\includegraphics[width=\textwidth]{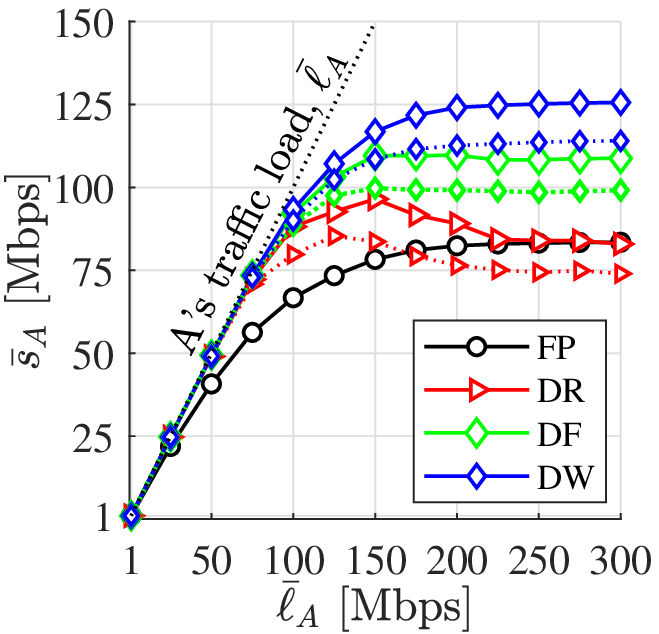}
			\caption{Average throughput.}
			\label{fig:throughput}
		\end{subfigure}
		\hfill
		\begin{subfigure}{0.24\textwidth}
			\includegraphics[width=\textwidth]{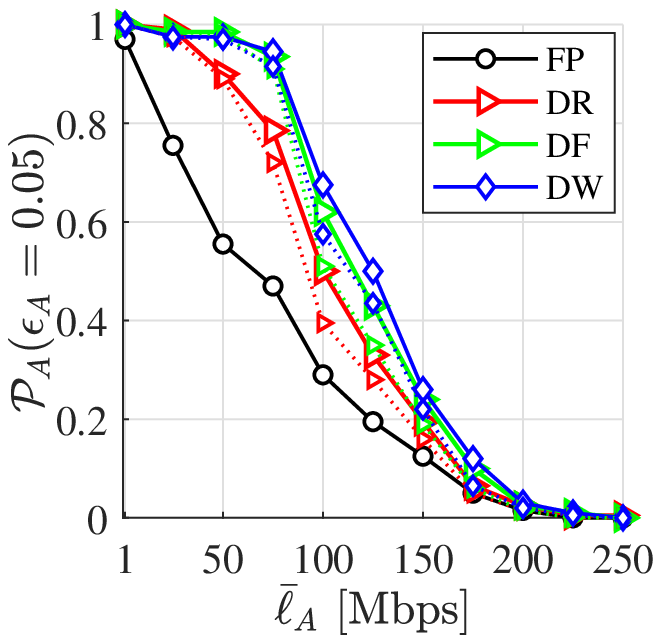}
			\caption{Prob. of \textit{desired throughput}.}
			\label{fig:prob_throughput}
		\end{subfigure}
		
		\caption{Performance metrics of WLAN A for the different primary channel selection algorithms. Continuous lines assume no switching cost while dashed lines correspond to $\delta=100$ ms.}
		\label{fig:results_central_scenario}
	\end{figure}
	
	\begin{figure}
		\includegraphics[width=0.48\textwidth]{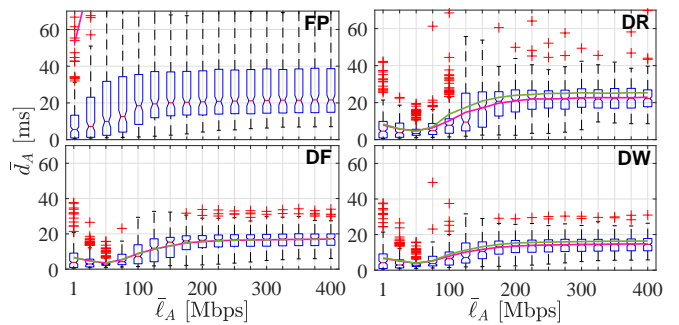}
		\caption{Distribution of the average delay of A for $\delta=0$ ms. The mean value for the different loads is represented through the purple and green lines for $\delta=0$ and $\delta=100$ ms, respectively.}
		\label{fig:delay}
	\end{figure}
	
	Figs. \ref{fig:results_central_scenario} and \ref{fig:delay} collect key performance metrics of WLAN A in the proposed scenario for different traffic loads. Namely, Fig. \ref{fig:throughput} shows the average throughput $\bar{s}_\text{A}$ computed as the number of bits successfully transmitted (acknowledged) divided by the simulation time. The probability $\mathcal{P}_\text{A}$ of successfully transmitting \textit{sufficient} traffic is plotted at Fig.~\ref{fig:prob_throughput}. Such a probability is computed as the portion of scenarios accomplishing $\bar{s}_\text{A} \geq (1 \Minus \epsilon_s) \bar{\ell}_\text{A}$, where $\epsilon_s = 0.05$ is set to deal with the non-deterministic traffic generation. Finally, Fig.~\ref{fig:delay} shows the average packet delay $\bar{d}_\text{A}$.
	
	As shown by the different metrics, DyWi clearly outperforms FP and DR both in terms of throughput and delay. In fact, even when considering a huge adaptation cost delay $\delta = 100$ \si{ms} (i.e., an important 10\% penalty with respect to the iteration duration $T$), $\mathcal{P}_A$ is prominently increased for moderate and medium loads.
	While DR may be counterproductive for high loads (see Fig. \ref{fig:throughput}), both DF and DyWi keep a constant performance after saturating. Such a throughput reduction for DR is caused by the fact that the larger $\bar{\ell}_A$, the harder to remain satisfied, which leads to more frequent channel switching. Then, critically for DR, the random selection of the primary leads to picking each channel with the same probability. Accordingly, the average throughput converges to FP's because, on average (for all the scenarios), the primary channels are equiprobable selected whenever the satisfaction condition is not accomplished.
	
	As for the average delay, even though we can see by the outliers in Fig. \ref{fig:delay} that adopting \eqref{eq:argmax} does not guarantee optimal performance for every scenario, DyWi clearly outperforms the rest algorithms in most of the cases.
	A phenomenon worth noticing is the reduction of the average delay from low to moderate traffic loads for online selection algorithms. During the normal operation of an unsatisfactory iteration, while for low traffic loads, the number of aggregated packets per frame is pretty low, 
	it is much greater for higher loads.
	Accordingly, since the buffer tends to remain unsaturated for low loads, during the new configuration setup after an unsatisfactory iteration, the buffer is normally significantly filled up. As a result, the first frame of the next iteration -- usually containing many aggregated data packets-- affects much more to the average delay for low traffic loads 
	than for high loads.
	
\begin{figure}[t]
		\centering
		\includegraphics[width=0.48\textwidth]{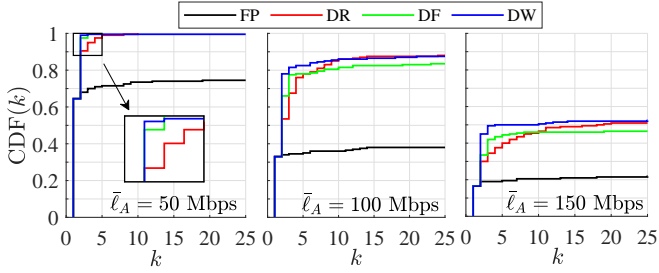}
		\caption{Cumulative distribution function of the number of iterations $k$ required to reach satisfaction.}
		\label{fig:cdf_first_happy_iteration}
	\end{figure}

	In order to assess the temporal evolution of the different algorithms, Fig. \ref{fig:cdf_first_happy_iteration} plots the cumulative distribution function (CDF) of the number of iterations $k$ required to reach a satisfactory primary channel for moderate, medium and high traffic loads. As expected, the lower the load, the higher the value of CDF($k$) for any $k$. Note that there are few unusual scenarios where the CDF varies for FP. Those are the cases where the load is almost adequate from the very beginning and the stochastic nature of the traffic generation makes the throughput to vary around $\eta\bar{\ell}_A$ as the simulation progresses.	
	
	It is also expected that the highest value of the CDF is normally provided by DyWi, which wisely adapts to the medium. However, DR may be even better for intricate scenarios (high $k$) and medium/high loads. This suggests that for such difficult scenarios, relying on \eqref{eq:argmax} may not be optimal due to the unexpected and harming interactions generated at the moment of changing the primary.
	Nevertheless, even though these unusual cases leave room for further improvement, DyWi is an effective solution that may be adopted in off-the-self WLANs due to its light-weight complexity and direct improvement. 
	
	\section{Conclusions}
	
	In this work, we have formulated DyWi, a lightweight and decentralized online primary channel selection algorithm for WLANs. DyWi aims at maximizing the throughput by iteratively estimating the occupancy of the primary and secondary channels, thus boosting potential bonds. Results show significant improvements with respect to traditional fixed allocation, even under the assumption of high adaptation costs.
	
	\bibliographystyle{IEEEtran}
	\bibliography{bib}
	
\end{document}